# Implications for galaxy property estimation revealed by CO luminosity-FWHM relations in local star-forming galaxies

Yi-Han Wu,[1,2,3]★ Jun-Feng Wang,[1]★ Xiao-Hu Li,[2,3]★ Xue-Jian Jiang,[4] Chao-Wei Tsai,[5,6] Jing-Wen Wu,[7,5] Kun-Peng Shi,[1] Lin Zhu[1] and Wen-Yu Zhong[1]

[1]*Department of Astronomy, College of Physical Science and Technology, Xiamen University, Xiamen 361005, P. R. China*
[2]*Xinjiang Astronomical Observatory, Chinese Academy of Sciences, Urumqi, Xinjiang 830011, P. R. China*
[3]*Xinjiang Key Laboratory of Radio Astrophysics, 150 Science-Street, Urumqi, Xinjiang 830011, P. R. China*
[4]*Research Center for Astronomical Computing, Zhejiang Laboratory, Hangzhou 311100, P. R. China*
[5]*National Astronomical Observatories, Chinese Academy of Sciences, 20A Datun Road, Beijing 100101, P. R. China*
[6]*Institute for Frontiers in Astronomy and Astrophysics, Beijing Normal University, Beijing 102206, P. R. China*
[7]*University of Chinese Academy of Sciences, Beijing 100049, P. R. China*

## ABSTRACT

This study explores a relationship between the CO luminosity-full width at half-maximum linewidth linear relation (i.e. the CO LFR) and mean galaxy property of the local star-forming galaxy sample in the xCOLDGASS data base, via a mathematical statement. The whole data base galaxies were separated into two subsamples based on their stellar masses and redshifts, being a help to examine the dependence issue of the CO LFR. Selecting the galaxy data with a stringent requirement was also implemented in order to assure the validly of the CO LFR. An algorithm of the linear regression was conducted with the data of the subsample. An assessment about the linear correlation manifested a valid CO LFR occurs in the selected galaxy of the subsample, and the intercept of the CO LFR may be related with the mean galaxy properties such as the molecular gas fraction and galaxy size. For the finding on the intercept of the CO LFR, we aligned that intercept with those galaxy properties via the involvement of a $\psi$ parameter. On evaluating the $\psi$ value with our local star-forming galaxy sample, we numerically determined a relationship between the statistical result and the galaxy property in a different stellar mass range. It also shows a possible method on estimating galaxy property.

## 1 INTRODUCTION

On estimating the properties (e.g. the mass, dynamical speed, luminosity, and morphological feature) of a galaxy, measuring the electromagnetic radiation generated from the atom and molecule within the galaxy had been a manner, and some statistical relationship or method useful on the study of galaxy followed that manner.

The radiations generated from atomic hydrogen (H) and molecular hydrogen ($H_2$) had been detected as tracers to reveal the luminosity and dynamical speed of galaxy (e.g. Kerr, Hindman & Carpenter 1956; Fish 1961; Roberts 1962; Höglund & Roberts 1965), and, there were scaling relations proposed based on those two measured quantities. For instance, the luminosity of an elliptical galaxy was found to correlate with its velocity dispersion positively (named by Faber–Jackson Relation which is shortly denoted as FJR in short; Faber & Jackson 1976), additionally, the luminosity of a spiral galaxy had been found to positively correlate with its rotational velocity, (named as Tully–Fisher Relation which is shortly denoted as TFR; Tully & Fisher 1977).

Apart from the radiations of H and $H_2$, carbon monoxide (CO[1]), a second abundance while comparing to that of $H_2$, was oftenly taken to trace the property (e.g. the mass, velocity, luminosity, and mass density) of galaxy (e.g. Young et al. 1995; Shull, Penton & Stocke 1999; Donahue et al. 2000; Carilli, Gnedin & Owen 2002; Greve, Ivison & Papadopoulos 2004; Solomon & Bout 2005; Harris et al. 2010, 2012; Emonts et al. 2011; Bothwell et al. 2013; Carilli & Walter 2013; Zanella et al. 2018; Salak et al. 2019; Le Fèvre et al. 2020; Neri et al. 2020). The CO flux was detected due to its rotational-transition emission[2] excited by the variance of the galaxy temperature which was firmly relevant to the galactic mass and pressure. Therefore, while measuring with the CO emission profile as well as its flux, one was able to trace the galaxy property.

---

[1]The $^{12}C^{16}O$ is the most abundant variety in the galaxy observation, and, deriving from this most abundant variety, there are other derivative isotopic types such as $^{13}C^{16}O$, $^{12}C^{18}O$, and $^{13}C^{18}O$. Without an explicit indication, due to the $^{12}C^{16}O$ being the most abundant variety, conventionally, it is symbolized as CO

[2]The rotational-transition emission of CO is generated while the transition on its two rotational energy levels occurs. In quantum physics, to a diatomic molecule, its rotational energy level is symbolized with a letter of $J$. A sequence of positive integers (0, 1, 2, and so on) is assigned to the $J$ in order to indicate a specific rotational energy level of a diatomic molecule, i.e. $J = 0$ is indicated as the basic (ground) level, $J = 1$ the first, $J = 2$ the second, and so on. Thus, to a CO molecule, a presentation of CO($J = 1 \rightarrow 0$)

★ E-mails: yhwastroph@xmu.edu.cn (Y-HW); jfwang@xmu.edu.cn (J-FW); xiaohu.li@xao.ac.cn (XHL)

Among the CO-measured data of galaxy, a statistical relationship similar to the TFR and FJR was proposed: a correlation between the CO luminosity and the full width at half-maximum (FWHM) linewidth measured in the CO line profile (e.g. Ho 2007; Davis et al. 2011; Goto & Toft 2015; Tiley et al. 2016; Topal et al. 2018), which was conventionally named as the CO luminosity–FWHM relation. Hereafter in this paper, we name the CO luminosity–FWHM relation as CO LFR for convenience. The physical speed characterizing the whole galactic system was adopted both in verifying the TFR and FJR, however, the CO FWHM linewidth was straightforwardly adopted to the verification of the CO LFR, indicating a potential uncertainty caused by the transformation between the CO FWHM linewidth and the physical speed of a galaxy could be eschewed.

By that proposal of the CO LFR, it could be also applied on, for instance, measuring the cosmic distance of a galaxy (e.g. Goto & Toft 2015), estimating the magnification factor of a gravitational-lensed galaxy (e.g. Harris et al. 2012; Goto & Toft 2015; Neri et al. 2020), estimating the value of molecular gas fraction to a galaxy (e.g. Isbell, Xue & Fu 2018), and constraining the cosmic parameter to uncover the time-varying nature of dark energy (e.g. Wu et al. 2019). Among those instances, a prominent study proposed in Isbell et al. (2018) (hereafter denoted as I18) motivated us to further concern the connection between the CO LFR and mean property of a galaxy sample.

In I18, the CO LFR was used to estimate the mean value on molecular gas fraction ($f_{mol}$, the ratio of the $H_2$ mass to dynamical mass of a galaxy or a group of galaxies) to the galaxy sample compiled. By the definition, the $f_{mol}$ value indicates how much a hydrogen mass is depleted in a galaxy. Thus, by the presentation of the mean $f_{mol}$ result estimated for their galaxy sample locating in a wide redshift range, i.e. $0.03 < z < 3.26$, they concluded that the mean galactic molecular gas fraction increases at a more rapid rate along the redshift.

As seen in fig. 1 of I18, six galaxy populations were showing a trend of the normalized molecular gas fraction ($f'_{mol}$). In the same figure, among the six CO LFRs firmly established in these six galaxy populations, can their intercepts and slopes be related to the $f'_{mol}$ trends? Or, be general, as we suspect, there could be a link between the statistical property of the CO LFR on the un-normalized $f'_{mol}$ (i.e. the $f_{mol}$) and other property of the galaxy (e.g. the mass content).

Therefore, to confirm, as being distinctive to the study in I18, we seek and show a possible connection between the CO LFR property and physical parameter in a galaxy sample, and, being a priority, the correlation significance of a CO LFR should be verified. This paper is organized as follows: Section 2 describes the galaxy sample in our compilation, Section 3 describes the methods adopted for the data selection in the galaxy sample and analysis, Section 4 presents the result and discussion with a further discussion for the result, and Section 5 presents the conclusion and possible future work for this study.

## 2 ARCHIVAL DATA

In our compilation for the CO data of galaxies, we noticed that most of the well-measured CO data were available for the galaxies observed in the local Universe, i.e. $z < 1$. Especially, the xCOLDGASS survey project of the IRAM[3] (Saintonge et al. 2017) was a successful accomplishment for providing the CO data in the $J = 1 \rightarrow 0$ and the $J = 2 \rightarrow 1$ transitions toward the 532 galaxies locating in $0.01 < z < 0.05$. In that survey project, a fits-format data base named xCOLDGASS_PubCat.fits was also publicly released.

The data structure of the xCOLDGASS_PubCat.fits is constructed by 532 rows and 71 columns, and the each row represents the survey galaxy following with its information listed from the Column 1 to 71. In the same data base, the APEX CO($J = 2 \rightarrow 1$)-measured datum of the survey galaxy is listed from the Column 65 to 71. The whole data base of the xCOLDGASS_PubCat.fits consists of two galaxy groups: one group includes 166 galaxies observed in $0.01 < z < 0.02$ with their stellar masses ranging in $10^9 \, M_\odot < M_* < 10^{10} \, M_\odot$, and the other includes 366 galaxies observed in $0.025 < z < 0.05$ with stellar masses ranging in $10^{10} \, M_\odot < M_* < 10^{11.5} \, M_\odot$. In addition, a supplementary document named xCOLDGASS_README.rtf was also attached to describe the content of the each column in the data base of the xCOLDGASS_PubCat.fits.

## 3 METHODS

### 3.1 Galaxy selection process

We conducted a series of selections among the 532 galaxies of the xCOLDGASS data base. For being clear and simple, in the following texts, the boldfaced terms themselves are used to not only indicate the column titles in the xCOLDGASS data base (check the column description with the xCOLDGASS_README.rtf), but also represent the physical quantities involved in different sections of this paper.

According to the study of Goto & Toft (2015), the higher degree of the correlation significance occurred under the use of the CO($J = 1 \rightarrow 0$) data of galaxies. Thus, in our decision, the CO($J = 1 \rightarrow 0$) data in the xCOLDGASS data base was adopted for the investigation of the CO LFR.

As described in the xCOLDGASS_README.rtf, to the column of the **AGNCLASS**, it indicates the whole xCOLDGASS galaxies are multiple in type, and they are classified with six indices, i.e. the Undetermined by the index of the $-1$, the Inactive by the 0, the SFing (star-forming galaxy) by the 1, the Composite by the 2, the active galactic nuclei (AGN) by the 3, and the Seyfert by the 4, not to mention the in-homogeneity in their data. To reduce the influence of a possible uncertainty caused by the different galaxy type and in-homogeneous data on our subsequent CO LFR investigation, we required that the galaxy sample as well as its datum involved in the CO LFR investigation should be similarly homogeneous in type and value, respectively. Some criteria on our galaxy selection were taken, and they are stated in the following three subsections.

#### 3.1.1 The first step: the homoegeneousness and the data availability

We required the galaxy whose type was being in the SFing (the index of the 1) since the domination in number occurred at this type. Also, we required the galaxy whose CO($J = 1 \rightarrow 0$) emission had been detected significantly, i.e. **FLAG_CO** = 1 and **SN_CO** > 5, adding its inclination (the **INCL**) and FWHM linewidth (the **WCO_TFR**) had been certainly measured both (**WCO_TFR** ≠ 0 and **INCL** ≠ 0). Thus, there were 125 galaxies remained by those requirements.

---

is taken to indicate that a transition on its $J$ occurs from the first level ($J = 1$) to the basic one ($J = 0$), and a corresponding radiation is emitted due to this transition. That transition presentation is conventionally adopted in astronomy, and adopted in this article.

[3]The Institut de radioastronomie millimétrique, whose website of https://iram-institute.org is public.

### 3.1.2 The second step: the sample separation

Recalling the data structure mentioned in Section 2, some of the galaxies were observed in $0.01 < z < 0.02$ with their stellar ranging in $10^9 \, M_\odot < M_* < 10^{10} \, M_\odot$ and the others in $0.025 < z < 0.05$ with their stellar masses ranging in $10^{10} \, M_\odot < M_* < 10^{11.5} \, M_\odot$. In following the first step, while checking the remained 125 galaxies, two galaxies, the 108 045 and 109 028, were both measured in the lower range of $0.01 < z < 0.02$, however, strangely, with their stellar masses ranging in the higher range of $10^{10} \, M_\odot < M_* < 10^{11.5} \, M_\odot$. Since that, we excluded those two galaxies from the 125, and hence 123 galaxies were being remained. Then, the 123 galaxies were separated into two subsamples as the xCOLDGASS-low and xCOLDGASS-high. In clarity, there were 46 galaxies allocated as the xCOLDGASS-low ($0.01 < z < 0.02$ and $10^9 \, M_\odot < M_* < 10^{10} \, M_\odot$) and 77 ones as the xCOLDGASS-high ($0.025 < z < 0.05$ and $10^{10} \, M_\odot < M_* < 10^{11.5} \, M_\odot$).

With that separation, it could be allowed us to ascertain whether the statistical outcome from the CO LFR investigation was dependent on the redshift and/or stellar mass of the subsample galaxy. For each galaxy of the two subsamples, its data on the **LCO_COR** (K km $s^{-1}$ pc²), **LCO_COR_ERR** (K km $s^{-1}$ pc²),[4] **WCO_TFR** (km $s^{-1}$), and **WCO_TFR_ERR** (km $s^{-1}$) were all certainly given.

### 3.1.3 The third step: the data processing

Processing the datum to our subsample galaxy before the investigation was necessarily considered. It should notice that the datum in the **WCO_TFR** was a quantity projected along the light of sight due to the inclination of the observed galaxy. Therefore, a correction for the galaxy inclination had to be considered in order to de-project the **WCO_TFR** datum. By the datum of the **INCL** (in the unit of degree, with no corresponding error provided in the xCOLDGASS data base), a projected **WCO_TFR** datum could be de-projected via the method (e.g. Tiley et al. 2016; Topal et al. 2018) expressed as:

$$W_{\text{corr}} = \frac{\text{WCO\_TFR}}{\sin(\text{INCL})} \,, \tag{1}$$

where the $W_{\text{corr.}}$[5] represents the outcome after the inclination correction with the **WCO_TFR** datum. Correspondingly, the measurement error of the $W_{\text{corr.}}$ can be obtained by

$$eW_{\text{corr}} = \frac{\text{WCO\_TFR\_ERR}}{\sin(\text{INCL})} \,, \tag{2}$$

where the $eW_{\text{corr}}$ represents the measurement error of the $W_{\text{corr.}}$. To each galaxy of the two subsamples, its **WCO_TFR** and **WCO_TFR_ERR** data were respectively converted into the $W_{\text{corr}}$ and $eW_{\text{corr}}$.

For the galaxies in the xCOLDGASS-low, while checking the statistical distribution on their **LCO_COR** data, the mean ($m_{(\text{LCO\_COR})}$) and standard deviation (SD, $\sigma_{(\text{LCO\_COR})}$) from their **LCO_COR** data were indicating that some of the **LCO_COR** data deviate from the mean value over around the 2 $\sigma_{(\text{LCO\_COR})}$ boundary. Similarly, in the same subsample, as checking the distribution on their $W_{\text{corr}}$ data, the mean ($m_{(W_{\text{corr}})}$) and SD ($\sigma_{(W_{\text{corr}})}$) from the $W_{\text{corr}}$ data were also indicating that some of the data deviate from the $m_{(W_{\text{corr}})}$ over the 2 $\sigma_{(W_{\text{corr}})}$. In the xCOLDGASS-high, by checking the statistical distributions on its **LCO_COR** and $W_{\text{corr}}$, we also noticed their deviation signs.

Here was a consideration that, statistically, the use of a high deviation data may cause a high systematic uncertainty occurring in the CO LFR investigation. Thus, to the galaxies in the xCOLDGASS-low, two rules,

$$m_{(\text{LCO\_COR})} - 2*\sigma_{(\text{LCO\_COR})} \qquad < \text{LCO\_COR}$$
$$< m_{(\text{LCO\_COR})} + 2*\sigma_{(\text{LCO\_COR})} \tag{3}$$

and

$$m_{(W_{\text{corr}})} - 2*\sigma_{(W_{\text{corr}})} < W_{\text{corr}} < m_{(W_{\text{corr}})} + 2*\sigma_{(W_{\text{corr}})} \tag{4}$$

, were both adopted to exclude the data which more deviate from its mean. The formats in equations (3) and (4) were also as the rules adopted to the galaxy data in the xCOLDGASS-high.

Actually, in our test, a strict requirement with the one time of the SD left the subsample galaxy in a less number, indicating an unreliable outcome in the CO LFR investigation. Thus, as the common requirement to the both subsamples, the manner of the two times of the SD had been a more suitable assignment in the data exclusions.

We also considered the influences of the uncertainties of the **LCO_COR** and $W_{\text{corr}}$ on the correlation significance of the CO LFR. Owing to the utilization of a log–log space in the CO LFR investigation, the **LCO_COR** and $W_{\text{corr}}$ data would be logarithmically changed, i.e. the $\log_{10}(\text{LCO\_COR})$ and $\log_{10}(W_{\text{corr}})$. Correspondingly, the uncertainties of the **LCO_COR** and $W_{\text{corr}}$ data should be respectively changed by the forms of

$$e\log_{10}(\text{LCO\_COR}) = \frac{1}{\text{Ln}(10)} \times \left( \frac{\text{LCO\_COR\_ERR}}{\text{LCO\_COR}} \right) \tag{5}$$

and

$$e\log_{10}(W_{\text{corr}}) = \frac{1}{\text{Ln}(10)} \times \left( \frac{eW_{\text{corr}}}{W_{\text{corr}}} \right) \,, \tag{6}$$

where the $e\log_{10}(\text{LCO\_COR})$ and $e\log_{10}(W_{\text{corr}})$ denote the logarithmic uncertainties of the $\log_{10}(\text{LCO\_COR})$ and $\log_{10}(W_{\text{corr.}})$. Moreover, since the uncertainty on the **INCL** was not involved in the calculation on $eW_{\text{corr}}$, the term $\frac{eW_{\text{corr}}}{W_{\text{corr}}}$ in equation (6) can be straightforwardly equal as $\frac{\text{WCO\_TFR\_ERR}}{\text{WCO\_TFR}}$, thus,

$$e\log_{10}(W_{\text{corr}}) = \frac{1}{\text{Ln}(10)} \times \left( \frac{\text{WCO\_TFR\_ERR}}{\text{WCO\_TFR}} \right) \,. \tag{7}$$

As shown in both equations (5) and (7), the term of $\frac{1}{\text{Ln}(10)}$ was as a common factor to correct the ratio of $\frac{\text{LCO\_COR\_ERR}}{\text{LCO\_COR}}$ as well as of $\frac{\text{WCO\_TFR\_ERR}}{\text{WCO\_TFR}}$. Then, obviously, the $e\log_{10}(\text{LCO\_COR})$ value was merely dependent on the $e\log_{10}(\text{LCO\_COR})$ value, and so was the $e\log_{10}(W_{\text{corr}})$ value.

In the xCOLDGASS-low galaxies, as checking the statistical distributions on their $\frac{\text{LCO\_COR\_ERR}}{\text{LCO\_COR}}$ and $e\log_{10}(\text{LCO\_COR})$, we found that the $e\log_{10}(\text{LCO\_COR})$ values were confined below 10 per cent when the $\frac{\text{LCO\_COR\_ERR}}{\text{LCO\_COR}}$ values were being confined below the $\frac{\text{LCO\_COR\_ERR}}{\text{LCO\_COR}}$ mean ($m_{\left(\frac{\text{LCO\_COR\_ERR}}{\text{LCO\_COR}}\right)}$). Similarly, to the same xCOLDGASS-low galaxies, we also found the below-10 per cent sign by checking the statistical distributions on the $\frac{\text{WCO\_TFR\_ERR}}{\text{WCO\_TFR}}$ and $e\log_{10}(W_{\text{corr}})$. Through the same checking-statistical-distribution manners implemented to the xCOLDGASS-low galaxies, we also found the below-10 per cent signs on the $\frac{LCO\_COR\_ERR}{LCO\_COR}$ and $\frac{WCO\_TFR\_ERR}{WCO\_TFR}$ data of the xCOLDGASS-high galaxies.

---

[4] According to the statement in Saintonge et al. (2017), both the **LCO_COR** and **LCO_COR_ERR** data had been calculated based on a set of cosmology parameters (i.e. $H_0 = 70$ km $s^{-1}$ Mpc$^{-1}$, $\Omega_m = 0.3$, and $\Omega_\Lambda = 0.7$) and corrected for the aperture issue. Thus, as our decision, they were adopted in this study.

[5] The $W_{\text{corr.}}$ itself is not one of the identifiers in the xCOLDGASS_PubCat.fits. The value of the $W_{\text{corr.}}$ was obtained by the process of our inclination correction.

**Table 1.** The galaxies selected in the subsample of the xCOLDGASS-low with their data used in this study. There are 18 galaxies eventually selected through Sections 3.1.1, 3.1.2, and 3.1.3. Based on the digital format in the xCOLDSGASS data base, the formats of the data listed in this table are presented in a simpler and more complete manner. For each galaxy in this table, the first column presents the identification (ID) number, the second presents the redshift, the third presents the inclination since unit is in degree, the fourth and fifth columns, respectively, present the inclination-corrected CO($J = 1{\to}0$) FWHM linewidth ($W_{corr}$, in km s$^{-1}$) and its measurement error ($eW_{corr}$, in km s$^{-1}$), the sixth and seventh columns, respectively, present the CO($J = 1{\to}0$) luminosity (the **LCO_COR**, in K km s$^{-1}$ pc$^2$) and its measurement error (the **LCO_COR_ERR**, in K km s$^{-1}$ pc$^2$), the eighth column presents the stellar mass ($M_*$, in M$_\odot$) in logarithm (the **LOGMSTAR**), the ninth column presents the molecular hydrogen mass ($M_{H_2}$, in M$_\odot$) in logarithm (the **LOGMH2**), the tenth column presents the $M_{H_2}$-to-$M_*$ ratio in logarithm (the **LOGMH2MS**), and the eleventh column presents the SDSS $r$-band major-axis effective radius (the **R50KPC**, in kpc).

| Galaxy ID | Redshift | INCL (Degree) | $W_{corr}$ (km s$^{-1}$) | $eW_{corr}$ (km s$^{-1}$) | LCO_COR (K km s$^{-1}$ pc$^2$) | LCO_COR_ERR (K km s$^{-1}$ pc$^2$) | LOGMSTAR $\log_{10}(\frac{M_*}{M_\odot})$ | LOGMH2 $\log_{10}(\frac{M_{H_2}}{M_\odot})$ | LOGMH2MS $\log_{10}(\frac{M_{H_2}}{M_*})$ | R50KPC (kpc) |
|---|---|---|---|---|---|---|---|---|---|---|
| 109010 | 0.010 19 | 54.5 | 263.65 | 33.79 | 117 524 432 | 22 874 044 | 9.75 | 8.75 | −1.00 | 2.20 |
| 108050 | 0.016 18 | 48.4 | 273.29 | 11.93 | 141 523 760 | 27 454 626 | 9.75 | 8.62 | −1.13 | 1.72 |
| 114041 | 0.012 93 | 33.7 | 219.02 | 6.32 | 229 606 128 | 39 826 372 | 9.84 | 8.88 | −0.96 | 1.74 |
| 112035 | 0.018 16 | 29.8 | 147.42 | 18.53 | 200 148 080 | 36 591 492 | 9.67 | 8.77 | −0.90 | 2.22 |
| 109038 | 0.011 54 | 46.4 | 205.24 | 15.54 | 29 547 702 | 6169 936 | 9.21 | 8.61 | −0.60 | 0.89 |
| 112050 | 0.014 20 | 56.8 | 109.00 | 8.06 | 44 963 744 | 8904 279 | 9.42 | 8.69 | −0.73 | 0.94 |
| 113067 | 0.018 75 | 32.5 | 235.25 | 24.87 | 228 519 888 | 46 120 748 | 9.89 | 8.89 | −1.00 | 3.86 |
| 109058 | 0.016 13 | 69.4 | 226.13 | 20.75 | 129 245 272 | 25 672 208 | 9.91 | 8.68 | −1.22 | 2.22 |
| 110038 | 0.016 67 | 49.3 | 288.41 | 49.55 | 119 189 040 | 23 592 062 | 10.00 | 8.67 | −1.33 | 1.31 |
| 109072 | 0.018 85 | 73.6 | 166.88 | 31.42 | 96 822 472 | 20 299 290 | 9.68 | 8.47 | −1.21 | 2.22 |
| 108093 | 0.017 76 | 55.4 | 210.40 | 23.80 | 228 745 024 | 42 048 240 | 9.94 | 8.85 | −1.09 | 1.69 |
| 123006 | 0.015 40 | 24.4 | 309.24 | 23.54 | 181 005 472 | 32 471 314 | 9.84 | 8.80 | −1.04 | 2.44 |
| 109106 | 0.016 28 | 52.5 | 149.54 | 20.26 | 132 582 688 | 25 697 748 | 9.65 | 8.75 | −0.90 | 2.57 |
| 114115 | 0.017 54 | 77.9 | 165.70 | 29.79 | 102 019 680 | 20 097 814 | 9.65 | 8.72 | −0.93 | 2.66 |
| 114121 | 0.014 89 | 69.0 | 222.99 | 26.08 | 121 559 528 | 23 776 134 | 9.77 | 8.76 | −1.02 | 2.46 |
| 113122 | 0.016 23 | 47.9 | 290.89 | 11.28 | 204 912 048 | 41 557 424 | 9.89 | 8.77 | −1.12 | 1.22 |
| 113136 | 0.016 63 | 69.8 | 130.70 | 23.80 | 104 160 088 | 21 504 220 | 9.78 | 8.75 | −1.03 | 3.84 |
| 113150 | 0.014 65 | 79.9 | 232.50 | 21.55 | 143 034 416 | 27 657 434 | 9.93 | 8.84 | −1.09 | 3.03 |

Following the above checkings, it was our consideration that both the elog$_{10}$(**LCO_COR**) and elog$_{10}$(**WCO_TFR**) values under the 10 per cent may cause a reduction on the systematic uncertainty of the CO LFR. Therefore, as a requirement of our galaxy selection, we confined the uncertainties of the **LCO_COR** and $W_{corr}$. To the xCOLDGASS-low galaxy, its ratios on the $\frac{LCO\_COR\_ERR}{LCO\_COR}$ and $\frac{WCO\_TFR\_ERR}{WCO\_TFR}$, respectively with their means (i.e. the $m_{(\frac{LCO\_COR\_ERR}{LCO\_COR})}$ and $m_{(\frac{WCO\_TFR\_ERR}{WCO\_TFR})}$), were adopted into two rules formed as

$$\frac{LCO\_COR\_ERR}{LCO\_COR} < m_{(\frac{LCO\_COR\_ERR}{LCO\_COR})} \quad (8)$$

and

$$\frac{WCO\_TFR\_ERR}{WCO\_TFR} < m_{(\frac{WCO\_TFR\_ERR}{WCO\_TFR})} \quad (9)$$

to select the $\frac{LCO\_COR\_ERR}{LCO\_COR}$ and $\frac{WCO\_TFR\_ERR}{WCO\_TFR}$ values which make the elog$_{10}$(**LCO_COR**) and elog$_{10}$(**WCO_TFR**) values both below 10 per cent.

The form in equations (8) and (9) were also as the rules to select the data in the xCOLDGASS-high.

Eventually, through the three steps of Sections 3.1.1, 3.1.2, and 3.1.3, there were 18 galaxies selected in the xCOLDGASS-low and 35 in the xCOLDGASS-high. Also, those eventually selected galaxies in the xCOLDGASS-low and xCOLDGASS-high, with their data used in this study, are respectively listed in Tables 1 and 2. Hereafter, in the following sections, both the xCOLDGASS-low and xCOLDGASS-high are meant as the subsamples after the three steps of the galaxy selection if an additional statement is indicating.

Next, we conducted the linear correlation analysis with the $L'_{CO(J=1{\to}0)}$ and FWHM$_{CO(J=1{\to}0)}$ (i.e. the **LCO_COR** and $W_{corr}$) data in the subsample to investigate whether the existence of the CO LFR is valid.

### 3.2 The linear regression analysis

Based on the plot in fig. 1 of I18, a log–log diagram was adopted by assigning the $\log_{10}$(**LCO_COR**) value as the dependent variable in the ordinate (i.e. the $y$-axis) and the $\log_{10}(W_{corr})$ value as the independent variable in the abscissa (i.e. the $x$-axis). It defined the plot frame for our linear correlation analysis. Then, as a conventional method in the analysis of the linear regression (LR), with four purely numerical parameters of $\alpha$, $\beta$, $A$ and $B$, a linear model was assumed as

$$(\log_{10}(\mathbf{LCO\_COR}) - A) = \alpha + \beta \times (\log_{10}(W_{corr.}) - B), \quad (10)$$

where the $\alpha$ and $\beta$, respectively, represent the intercept and slope of this linear model, and the $A$ and $B$, respectively, represent the median values on the $\log_{10}$(**LCO_COR**) and $\log_{10}(W_{corr.})$ data. The $A$ and $B$ are paired together to indicate a pivot point which is located among the data point depicted in the log–log diagram.

By the linear model in equation (10), an LR algorithm was conducted with using a Python package called LINMIX[6] to the xCOLDGASS-low as well as to the xCOLDGASS-high. By the LR algorithm, the $\alpha$ and $\beta$ values, which were simultaneously being fitted best to the data of the subsample, were derived numerically.

## 4 RESULT AND DISCUSSION

As looking at Fig.1, a linear feature was obviously presented in the xCOLDGASS-low data point, and this linear feature also occurred in the xCOLDGASS-high one. Besides, these two linear features showed their climbing-up trends both, indicating, in statistics, no matter in the xCOLDGASS-low or xCOLDGASS-high data point, the $\log_{10}(\mathbf{LCO\_COR})$ value was positively correlating with the $\log_{10}(W_{corr})$ one.

---

[6]Kelly (2007) constructed a Python package whose LR algorithm takes the measurement errors of the independent and dependent variables into account.

**Table 2.** The galaxies selected in the subsample of the xCOLDGASS-high with their data used in this study. There are 35 galaxies eventually selected through Sections 3.1.1, 3.1.2, and 3.1.3. Based on the digital format in the xCOLDSGASS data base, the formats of the data listed in this table are presented in a simpler and more complete manner. In this table, the descriptions for the each columns are the same as ones in Table 1.

| Galaxy ID | Redshift | INCL [Degree] | $W_{corr}$ [km s$^{-1}$] | e$W_{corr}$ [km s$^{-1}$] | LCO_COR [K km s$^{-1}$ pc$^2$] | LCO_COR_ERR [K km s$^{-1}$ pc$^2$] | LOGMSTAR $\log_{10}(\frac{M_*}{M_\odot})$ | LOGMH2 $\log_{10}(\frac{M_{H_2}}{M_\odot})$ | LOGMH2MS $\log_{10}(\frac{M_{H_2}}{M_*})$ | R50KPC [kpc] |
|---|---|---|---|---|---|---|---|---|---|---|
| 3819 | 0.045 25 | 29.5 | 418.13 | 11.60 | 2170 262 270 | 380 173 056 | 10.67 | 9.87 | −0.80 | 2.44 |
| 3962 | 0.042 59 | 57.1 | 363.02 | 8.38 | 2658 701 310 | 458 967 904 | 10.90 | 9.90 | −1.00 | 4.43 |
| 25763 | 0.029 61 | 38.0 | 131.49 | 13.89 | 559 401 220 | 101 763 744 | 10.11 | 9.22 | −0.89 | 4.79 |
| 26221 | 0.031 58 | 64.9 | 505.59 | 5.62 | 2689 082 620 | 459 958 976 | 10.98 | 9.91 | −1.07 | 4.93 |
| 23120 | 0.033 79 | 62.9 | 402.94 | 23.93 | 1124 069 250 | 205 938 624 | 10.03 | 9.05 | −0.98 | 2.17 |
| 24183 | 0.042 83 | 39.0 | 483.93 | 10.19 | 1935 294 850 | 347 787 328 | 10.77 | 9.83 | −0.94 | 5.03 |
| 6506 | 0.048 64 | 53.5 | 547.10 | 10.31 | 2314 337 540 | 419 841 376 | 10.77 | 9.95 | −0.82 | 5.38 |
| 26822 | 0.037 57 | 60.9 | 466.76 | 41.49 | 1611 847 810 | 290 347 872 | 11.03 | 9.73 | −1.30 | 5.31 |
| 13624 | 0.029 64 | 47.9 | 359.84 | 4.14 | 728 856 380 | 132 705 080 | 10.31 | 9.34 | −0.97 | 3.83 |
| 7493 | 0.026 23 | 66.6 | 388.82 | 8.01 | 996 845 500 | 176 116 368 | 10.56 | 9.47 | −1.10 | 2.70 |
| 10885 | 0.027 57 | 38.1 | 326.18 | 15.97 | 425 914 784 | 80 441 640 | 10.04 | 9.11 | −0.94 | 2.35 |
| 11349 | 0.025 56 | 58.5 | 199.60 | 9.62 | 370 441 952 | 67 617 056 | 10.13 | 9.03 | −1.10 | 2.85 |
| 11408 | 0.025 95 | 55.0 | 56.88 | 5.81 | 300 529 504 | 54 174 460 | 10.05 | 8.95 | −1.10 | 3.61 |
| 11845 | 0.036 43 | 69.7 | 320.00 | 14.30 | 1246 265 340 | 220 357 264 | 10.60 | 9.61 | −1.00 | 3.84 |
| 9551 | 0.026 80 | 58.2 | 550.22 | 3.96 | 2151 608 580 | 369 591 872 | 10.91 | 9.90 | −1.01 | 2.97 |
| 42013 | 0.036 94 | 52.4 | 392.17 | 7.12 | 2092 407 940 | 367 115 296 | 10.77 | 9.89 | −0.88 | 2.89 |
| 23194 | 0.034 00 | 57.8 | 354.93 | 7.04 | 1666 951 810 | 289 654 688 | 10.59 | 9.22 | −1.37 | 3.52 |
| 11340 | 0.035 72 | 66.6 | 303.72 | 10.01 | 696 956 030 | 126 332 432 | 10.10 | 9.53 | −0.57 | 2.46 |
| 4045 | 0.026 43 | 27.9 | 124.06 | 3.01 | 1173 051 390 | 200 090 352 | 10.47 | 9.56 | −0.91 | 1.54 |
| 51276 | 0.029 58 | 64.7 | 299.26 | 14.03 | 624 982 590 | 111 017 936 | 10.37 | 9.32 | −1.05 | 3.58 |
| 14712 | 0.038 23 | 18.9 | 429.74 | 12.98 | 1519 324 420 | 265 347 408 | 10.55 | 9.73 | −0.82 | 5.13 |
| 18673 | 0.038 45 | 48.8 | 388.38 | 15.10 | 902 163 140 | 167 842 032 | 10.38 | 9.44 | −0.94 | 2.04 |
| 26598 | 0.025 12 | 45.7 | 356.09 | 18.39 | 509 847 072 | 93 032 448 | 10.47 | 9.16 | −1.30 | 5.15 |
| 22822 | 0.026 99 | 66.9 | 372.72 | 5.27 | 638 607 810 | 116 893 232 | 10.56 | 9.40 | −1.16 | 2.29 |
| 28365 | 0.032 15 | 27.9 | 273.26 | 6.76 | 1018 264 960 | 177 244 944 | 10.36 | 9.50 | −0.86 | 6.68 |
| 31775 | 0.041 06 | 51.7 | 510.16 | 19.19 | 1423 559 300 | 258 902 192 | 10.57 | 9.15 | −1.42 | 3.64 |
| 32568 | 0.042 42 | 60.3 | 389.75 | 9.55 | 1079 164 030 | 190 106 304 | 10.36 | 9.56 | −0.79 | 4.68 |
| 12533 | 0.033 12 | 23.6 | 166.04 | 15.66 | 658 715 260 | 118 455 104 | 10.09 | 9.32 | −0.77 | 4.30 |
| 30332 | 0.042 41 | 44.1 | 411.80 | 39.01 | 2103 824 770 | 374 922 272 | 10.49 | 9.82 | −0.67 | 5.62 |
| 44718 | 0.038 30 | 43.0 | 433.50 | 2.79 | 1539 447 420 | 271 570 912 | 10.45 | 9.70 | −0.75 | 4.14 |
| 44942 | 0.033 04 | 47.8 | 474.14 | 49.70 | 1264 994 430 | 229 383 712 | 10.44 | 9.61 | −0.84 | 4.12 |
| 1115 | 0.025 97 | 34.1 | 261.66 | 9.90 | 941 179 840 | 165 540 976 | 10.11 | 9.54 | −0.57 | 4.86 |
| 1137 | 0.040 13 | 27.2 | 167.38 | 16.43 | 603 187 580 | 109 694 808 | 10.28 | 9.27 | −1.01 | 5.96 |
| 24973 | 0.028 54 | 27.9 | 141.01 | 3.09 | 2123 997 570 | 362 683 232 | 10.61 | 9.33 | −1.29 | 3.07 |
| 1221 | 0.034 63 | 16.9 | 457.29 | 24.77 | 1242 431 620 | 216 776 032 | 10.20 | 9.62 | −0.58 | 2.49 |

The LR algorithm results presented in Table 3 numerically show the above two linear features. For the xCOLDGASS-low, the Spearman's coefficient is 0.4 with the corresponding *p*-value of 0.1, indicating that an intermediately positive correlation on the $\log_{10}(\mathbf{LCO\_COR})$ and $\log_{10}(W_{corr})$ data is being manifested. For the xCOLDGASS-high, the Spearman's coefficient is 0.65 with the *p*-value of $2.5 \times 10^{-5}$, indicating a strongly positive correlation between the $\log_{10}(\mathbf{LCO\_COR})$ and $\log_{10}(W_{corr})$ data. Also, Fig. 1 presents the LR algorithm result for the subsample. In this figure, the blue point represents the $\log_{10}(\mathbf{LCO\_COR})$-$\log_{10}(W_{corr})$-paired data point of the xCOLDGASS-low galaxy and the red point represents the one in the xCOLDGASS-high galaxy. Moreover, whichever for the data point in the xCOLDGASS-low or xCOLDGASS-high, its horizontal and vertical bars represent the logarithmic errors derived for the e$W_{corr}$ and $\mathbf{LCO\_COR\_ERR}$, respectively. The tinted straight segment represents the best-fitting correlation line with the shaded area indicating the $1\sigma$ confidential area around the best-fitting correlation line (the blue for the xCOLDGASS-low and the red for the xCOLDGASS-high).

As a notice on the format of the linear model in equation (10), the 'true' intercept should be defined as a term of $A + \alpha - \beta \times B$ and the 'true' slope as the $\beta$. Hence, we evaluated the 'true' intercept and slope to the subsample via its statistical result in Table 3. Fig. 2 also presents the 'true' intercept (in the left panel) and 'true' slope (in the right panel) of the subsample over redshift (by the same colour denotations used in Fig. 1, the blue for the xCOLDGASS-low and the red for the xCOLDGASS-high).

Thus, we may assure that the CO LFR is significantly verified in both the xCOLDGASS-low and xCOLDGASS-high subsamples due to their positive and robust Spearman's coefficients. And, the positive Spearman's coefficient, whatever derived in the xCOLDGASS-low or xCOLDGASS-high, manifests that the higher CO luminosity occurs in the higher CO FWHM linewidth. By the CO LFR trend in Fig. 1, as considering that the $\log_{10}(W_{corr})$ range in the xCOLDGASS-low is consistently similar with that in xCOLDGASS-high (since the $\log_{10}(W_{corr})$ range in the xCOLDGASS-low is entirely included and overlapped by that in the xCOLDGASS-high), we also found the $\log_{10}(\mathbf{LCO\_COR})$ range in the former is on average lower than that in the latter, which is consistent to the $M_*$ range assignment in Section 3.1.2, i.e. the $M_*$ range of the xCOLDGASS-low is entirely lower than that of the xCOLDGASS-high. In actual, that finding from the CO LFR trend is as well consistent to the evidence inferring in the extragalactic observation, i.e. the CO luminosity of an extragalaxy is proportional to the mass it contains.

**Table 3.** The subsample involved in the LR analysis and its corresponding result from the LR analysis. The first column presents the name of the subsample, and the parenthesis behind the name shows the galaxy number in this subsample. The second, third, and fourth columns, respectively, present the numerical results for the $\alpha$, $\beta$, and intrinsic scatter. The intrinsic scatter indicates the degree of the data point scattering around the best-fitting correlation line. The fifth column presents the Spearman's coefficient used to indicate the strength of the correlation between the $\log_{10}(\mathbf{LCO\_COR})$ and $\log_{10}(W_{\mathrm{corr}})$ data, following with the parenthesis which insides shows a corresponding $p$-value. A threshold is adopted to the $p$-value, and there is a showing of a significant $\log_{10}(\mathbf{LCO\_COR})$-$\log_{10}(W_{\mathrm{corr}})$ correlation if resulting $p$-value is smaller than the threshold (0.1, or 0.05 rigorously). The sixth column represents the numerical median value of the $\log_{10}(\mathbf{LCO\_COR})$ data, and the seventh column represents the numerical median value of the $\log_{10}(W_{\mathrm{corr}})$ data. The value behind the $\pm$sign represents the $1\sigma$ uncertainty of the measurement value.

| Sample ($N$) | $\alpha$ | $\beta$ | Intrinsic scatter | Spearman's coefficient ($p$-value) | $A$ | $B$ |
|---|---|---|---|---|---|---|
| xCOLDGASS-low (18) | $0.010 \pm 0.060$ | $0.910 \pm 0.550$ | 0.230 | 0.40(0.100) | 8.120 | 2.340 |
| xCOLDGASS-high (35) | $0.020 \pm 0.040$ | $0.680 \pm 0.180$ | 0.210 | $0.65(2.5 \times 10^{-5})$ | 9.070 | 2.570 |

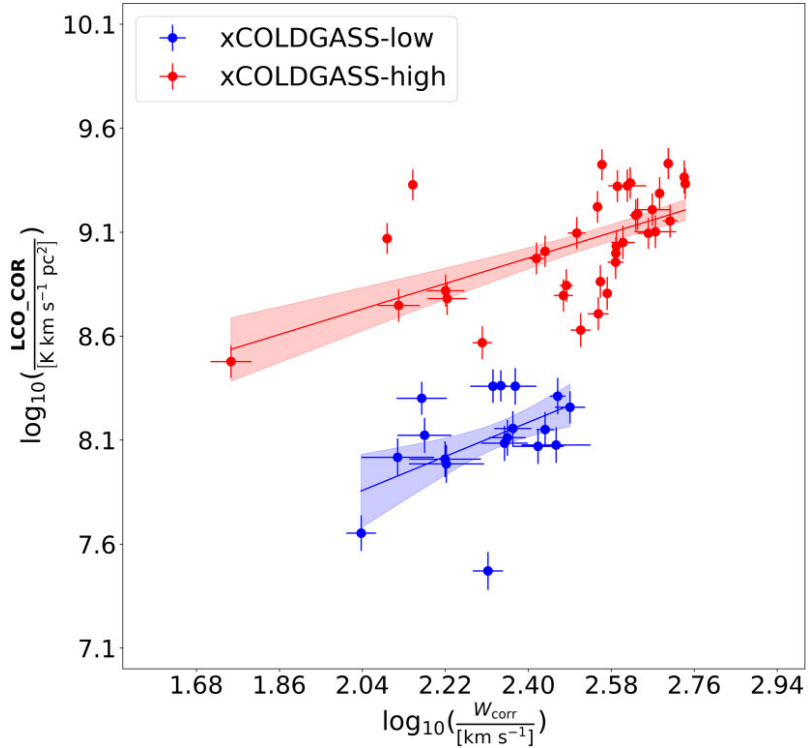

**Figure 1.** This figure displays the $\log_{10}(\mathbf{LCO\_COR})$–$\log_{10}(W_{\mathrm{corr}})$ data point of the subsample in the log–log diagram. The blue points represent the data in the xCOLDGASS-low and the red ones in the xCOLDGASS-high. In this figure, the linear segment represents the best-fitting linear correlation lines, with the shaded areas representing the $1\sigma$ error fields around the best-fitting lines, and the colour denotation used for the best-fitting line and shaded area follows the one used for that data point.

In Fig. 2, whether in the left (the intercept result) or right (the slope result) panel, the uncertainty (the $1\sigma$ errorbar) in the xCOLDGASS-low data point (the blue) is always larger than the one in the xCOLDGASS-high data point (the red). By noticing the intrinsic scatter results in Table 3, the intrinsic scatter in the xCOLDGASS-low is larger than the one in the xCOLDGASS-high. In statistics, the intrinsic scatter manifested by the data point of the subsample has an influence on the outcome of the property parameter, i.e. the higher the intrinsic scatter is, the larger the uncertainty of both the $\alpha$ and $\beta$ is. The degree of the intrinsic scatter was found to be negatively proportional to the size of the subsample, i.e. the larger intrinsic scatter result occurs in the subsample whose galaxy number is less, and it can be regarded as a reason causing the variety on intrinsic scatter. However, while checking the stringent requirements on our galaxy selection and the statistical non-negligibility on the galaxy numbers of the two subsamples, the inequality occurring in between the galaxy numbers could not be more dominant, with the correlation validity already confirmed by the robust value of the Spearman's Coefficient.

Again, in the right panel of Fig. 2, we found the slope measured in the xCOLDGASS-low is closely similar with the one in the xCOLDGASS-high. It indicates the slope of the best-fitting correlation line could be regarded as an independence of $M_*$ and $z$. However, in the left panel of Fig. 2, the intercept measured in the xCOLDGASS-high deviates from the one in the xCOLDGASS-low by a $2\sigma$ gap. Thus, the occurrence of that intercept deviation indicates the intercept itself may evolve with $M_*$ and $z$. While comparing the redshift ranges of our two subsamples to the current measured redshift from the observation, it can be regarded that there is no distinguishable difference in between these two subsample redshift ranges since, or actually, the galaxies in our subsamples were all observed in the local Universe. But, as looking in the $M_*$ ranges of our subsamples, i.e. for the xCOLDGASS-low, $10^9\,\mathrm{M_\odot} < M_* < 10^{10}\,\mathrm{M_\odot}$; for the xCOLDGASS-high, $10^{10}\,\mathrm{M_\odot} < M_* < 10^{11.5}\,\mathrm{M_\odot}$, a distinction on the $M_*$ makes this physics quantity becomes a dominating matter. Hence, we may infer that the occurrence of the $2\sigma$ intercept deviation indicates the intercept itself evolves with the $M_*$.

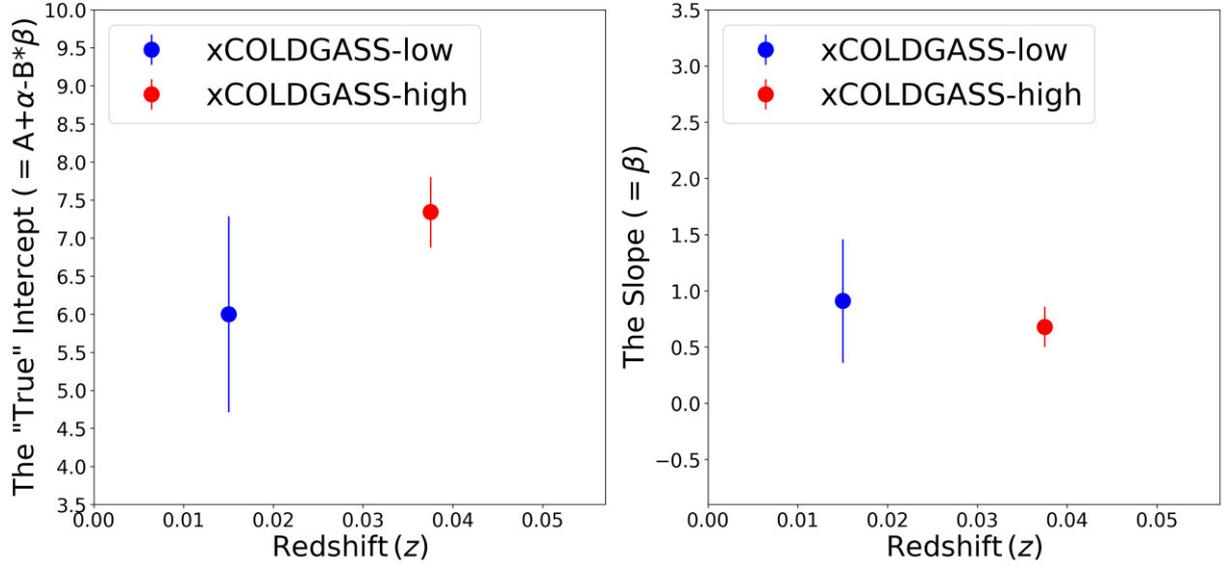

**Figure 2.** With the LR result in Table 3 and the colour denotation used in Fig. 1, in this figure, the LR result is depicted as a function of the median redshift of the subsample. The left panel presents the 'true' intercept result ($A + \alpha - \beta \times B$) with the median redshift, and the right panel presents the 'true' slope result ($\beta$) with the median one.

By recalling that the observed molecular gas fraction ($f_{mol}$) being dependent on the mass of galaxy, a further investigation in between the $f_{mol}$ and that inference on the $2\sigma$ intercept deviation would be started as another discussion.

### 4.1 Further investigation

To a galactic system (a galaxy or a group of galaxies), the $f_{mol}$ can be mathematically defined as a ratio between the molecular gas mass ($M_{mol}$, the total molecular hydrogen mass, $M_{H_2}$, including with the helium mass, $M_{He}$) and dynamical mass ($M_{dyn}$, an overall mass summing up the atomic hydrogen mass, $M_{HI}$, molecular hydrogen mass, $M_{H_2}$, and stellar mass, $M_*$),[7] that is,

$$f_{mol} = \frac{M_{mol}}{M_{dyn}} \ . \tag{11}$$

The $M_{mol}$ in equation (11) can be derived via

$$M_{mol} = \zeta \times L'_{CO} \ , \tag{12}$$

whose the $\zeta$[8] is presented as a conversion factor linking the $L'_{CO}$ (K km s$^{-1}$ pc$^2$) to the $M_{mol}$. A Greek letter of $\alpha$ had been conventionally adopted to link the $M_{mol}$ to the $L'_{CO}$ as shown in equation

[7]The units of the mass parameters (i.e. the $M_{mol}$, $M_{dyn}$, $M_*$, $M_{HI}$, $M_{H_2}$, and $M_{He}$) are all in M$_\odot$.
[8]In the unit of $\frac{\mathrm{M_\odot}}{\mathrm{K\ km\ s^{-1}\ pc^2}}$.

(12), however, the $\alpha$ was being used in equation (10). To avoid being confused, a Greek letter of $\zeta$ was taken to connect between the $M_{mol}$ and $L'_{CO}$. To the $M_{dyn}$ in equation (11), it can be measured from the dynamical rotational velocity $V_{rot}$ (km s$^{-1}$) characterizing a whole galactic system. That is, in the Newtonian Law, there is a relation linking the $M_{dyn}$ and $V_{rot}$ by

$$M_{dyn} = \frac{R \times V_{rot}^2}{G} \ , \tag{13}$$

where the $R$ (kpc) represents the characteristic radius of the galactic system[9] and the $G$ is the gravitational coefficient.[10] A scaling formula relating the physical velocity $V_{rot}$ to the measured FWHM$_{CO}$[11] can be simply assumed as

$$V_{rot} = C \times (\mathrm{FWHM_{CO}}) \tag{14}$$

via a dimensionless parameter of the $C$. Due to the relation in equation (14), the $V_{rot}$ in equation (13) can be replaced by the FWHM$_{CO}$, thus, equation (13) can be re-written as

$$M_{dyn} = \left( \frac{R \times C^2}{G} \right) \times (\mathrm{FWHM_{CO}})^2 \ . \tag{15}$$

For convenience, in equation (15), the form of $\frac{R \times C^2}{G}$ can be denoted by a letter of $k$, i.e. $\frac{R \times C^2}{G} = k$. Then, equation (15) can be re-written as

$$M_{dyn} = k \times (\mathrm{FWHM_{CO}})^2 \ . \tag{16}$$

[9]An arbitrary parameter should be presented as a factor to correct the characteristic radius $R$ while the physical feature of the galactic system is not in global. However, as our derivation in this subsection, to be simple, the factor correcting the issue due to the physical feature of the galactic system was already considered into the $R$ itself.
[10]Due to the units of the M$_\odot$, km s$^{-1}$, and kpc used in equation (13), correspondingly, the $G$ itself is equal as $4.3 \times 10^{-6}$ (kpc km$^2$ s$^{-2}$ M$_\odot^{-1}$).
[11]The FWHM$_{CO}$ is actually equal to the $W_{corr}$ defined in equation (1), however, we still use the denotation of the FWHM$_{CO}$ to represent a linewidth measured from a CO line profile, and the denotation of the $W_{corr}$ was used to represent a quantity adopted in the data base.

With the definitions expressed in equations (12) and (16), the $f_{gas}$ of equation (11) can be re-arranged as

$$f_{mol} = \frac{\zeta \times L'_{CO}}{k \times (\mathrm{FWHM_{CO}})^2} \, , \tag{17}$$

where the $\zeta$ and $k$, respectively, situated in the numerator and denominator can be re-arranged together and denoted by a letter of $\eta$, i.e. $\frac{\zeta}{k} = \eta$. Thus, the $f_{mol}$ in equation (17) can be simply formulated with the $\eta$ as

$$f_{mol} = \eta \times \frac{L'_{CO}}{(\mathrm{FWHM_{CO}})^2} \, , \tag{18}$$

which is presenting a formulation connecting among the $f_{mol}$, $L'_{CO}$, and $\mathrm{FWHM_{CO}}$.

Again, with equation (18), it can be arithmetically reformed into

$$\frac{f_{mol}}{\eta} \times (\mathrm{FWHM_{CO}})^2 = L'_{CO} \, . \tag{19}$$

Taking the logarithm on the left- and right-hand sides of equation (19), an expression can be derived as

$$\log_{10}(L'_{CO}) = \log_{10}\left(\frac{f_{mol}}{\eta}\right) + 2 \times \log_{10}(\mathrm{FWHM_{CO}}) \, . \tag{20}$$

Then, in equation (20), by assigning the $\log_{10}(\mathrm{FWHM_{CO}})$ and $\log_{10}(L'_{CO})$ as the independent and dependent variables, the whole expression in equation (20) can be regarded as a linear function with the $\log_{10}(\frac{f_{mol}}{\eta})$ as its intercept and the 2 as its slope.

Moreover, one should recall the linear model in equation (10), and that linear model is re-arranged as

$$\log_{10}(\mathbf{LCO\_COR}) = A + \alpha - \beta \times B + \beta \times \log_{10}(W_{corr}) \, . \tag{21}$$

While checking equations (20) and (21), commonly, their logarithmic CO luminosities [the $\log_{10}(L'_{CO})$ in equation 20 and the $\log_{10}(\mathbf{LCO\_COR})$ in equation 21] and logarithmic CO FWHM linewidths (the $\log_{10}(\mathrm{FWHM_{CO}})$ in equation 20 and the $\log_{10}(W_{corr})$ in equation 21) were respectively assigned as the dependent and independent variables. A skillful analogousness was implemented to equations (20) and (21). While looking in the slope portions of equations (20) and (21), one could regard the $\beta$ of equation (21) corresponds to a theoretical constant denoted by the 2 of equation (20), which may indicate that the slope of the statistical linear model may be in fix and inclination of the redshift and stellar mass of galaxy, as found in our LR results.

Again, via the same analogous technique to equations (20) and (21), in their intercept portions, the form of $\log_{10}(\frac{f_{mol}}{\eta})$ in equation (20) could be corresponded to the form of $A + \alpha - \beta \times B$ in equation (21). The two intercept potions can be proportional to each other by a proportional form of

$$A + \alpha - \beta \times B \propto \log_{10}\left(\frac{f_{mol}}{\eta}\right) \, . \tag{22}$$

Herein, in equation (22), it presents a relationship between the statistical behaviours (i.e. the $A$, $B$, $\alpha$, and $\beta$) of the linear model and the physics quantities (i.e. the $f_{mol}$ and $\eta$). Additionally, the $\eta$ itself is relevant to the $\zeta$ the type stellar and mass of galaxy, and the $f_{mol}$ itself was found to vary with the stellar mass of galaxy as well. Since those, in equation (18), that relation manifested that the 'true' intercept ($A + \alpha - \beta \times B$) of the correlation line is indeed sensitive to the variances on the $f_{mol}$ and $\eta$, as found in our LR result.

In further, to be pragmatic in use, we equalized the left- and right-hand sides of the proportional relationship in equation (22) through

involving an arbitrary parameter of $\psi$, which can be expressed as

$$A + \alpha - \beta \times B = \psi \times \log_{10}\left(\frac{f_{mol}}{\eta}\right) \, . \tag{23}$$

Moreover, recalling $\eta = \frac{\zeta \times G}{R \times C^2}$, the identity in equation (23) can also be expressed as

$$A + \alpha - \beta \times B = \psi \times \log_{10}\left(\frac{f_{mol} \times R \times C^2}{\zeta \times G}\right) \, . \tag{24}$$

The value of the $\psi$ in equation (24) was undetermined. As an initial test, to evaluate the value of the $\psi$ with our subsample, its mean values on the $f_{mol}$, $R$, $C$, and $\zeta$ and its results on the statistical CO LFR were utilized.

An algorithmic skill was applied to manifest the $\psi$ alone by formulating the $A + \alpha - \beta \times B$ and $\log_{10}\left(\frac{f_{mol} \times R \times C^2}{\zeta \times G}\right)$ parts of the identity in equation (24) as the numerator and denominator of a fraction. Thus, via that manner, the $\psi$ can be manifested as

$$\psi = \frac{A + \alpha - \beta \times B}{\log_{10}\left(\frac{f_{mol} \times R \times C^2}{\zeta \times G}\right)} \, . \tag{25}$$

In the formulation of equation (25), the $\alpha$ and $\beta$ parameters themselves are two purely numerical parameters assumed to scale between the logarithmic CO luminosity and logarithmic CO linewidth. To the dimensions of the $A$ and $B$ (see their sources in Section 3.2), as a technical consideration in this initial test, one could assume that both the units of the **LCO_COR** and $W_{corr}$ had been normalized into one after being processed by the manners of the median and logarithm, which similarly could do so to the part of $\log_{10}\left(\frac{f_{mol} \times R \times C^2}{\zeta \times G}\right)$. Thus, by that assumption on the parameter dimension, we let the $\psi$ to be a purely numerical factor scaling the numerator ($A + \alpha - \beta \times B$) and denominator ($\log_{10}\left(\frac{f_{mol} \times R \times C^2}{\zeta \times G}\right)$).

After that brief dimensional analysis, the evaluation for the $\psi$ value can be started. The $f_{mol}$ is defined as:

$$f_{mol} = \frac{M_{H_2}}{M_{H_2} + M_{HII} + M_*} \, , \tag{26}$$

in term of the $M_{HI}$, $M_{H_2}$, and $M_*$. In the right-hand side of the equal mark in equation (26), both the numerator and denominator portions were multiplied with a term of $\frac{1}{M_*}$, and then, the mass fraction form in equation (26) could be algorithmically changed as

$$f_{mol} = \frac{\left(\frac{M_{H_2}}{M_*}\right)}{\left(\frac{M_{H_2}}{M_*} + \frac{M_{HII}}{M_*} + 1\right)} \, . \tag{27}$$

According to the subsection 2.3 in Saintonge et al. (2017), the $\frac{M_{HI}}{M_*}$ gas fraction of the galaxy sample in the low-mass extension was restricted in between 2% and 10%. Thus, due to that restriction information, for our xCOLDGASS-low, we assumed a percent-formated value of 6% (= $\frac{(0.10+0.02)}{2}$) to the $\frac{M_{HI}}{M_*}$. Besides, in our xCOLDGASS-low, the mean value of $\frac{M_{H_2}}{M_*}$ was 10% with the SD of 0.05 based on the tenth column of Table 1. Again, in the subsection 2.3 of Saintonge et al. (2017), the $\frac{M_{HI}}{M_*}$ gas fraction of the high-mass galaxy was restricted in between 2% and 5%. Hence, for our xCOLDGASS-high, we assumed a value of 3.5% (= $\frac{(0.05+0.02)}{2}$) to the $\frac{M_{HI}}{M_*}$, and, based on the tenth column of Table 2, in our xCOLDGASS-high, the mean value of $\frac{M_{H_2}}{M_*}$ was 12% with the SD value of 0.06. By those assumptions to the $\frac{M_{HI}}{M_*}$ and $\frac{M_{H_2}}{M_*}$, we were able to determine the values on the $f_{mol}$ via equation (27) to the two subsamples, which can be summarized as, for the xCOLDGASS-low,

$$f_{mol} = \frac{\left(\frac{M_{H_2}}{M_*}\right)}{\left(\frac{M_{H_2}}{M_*} + \frac{M_{HI}}{M_*} + 1\right)} = \frac{10\%}{10\% + 6\% + 1} = 8.6\% , \quad (28)$$

and, for the xCOLDGASS-high,

$$f_{mol} = \frac{\left(\frac{M_{H_2}}{M_*}\right)}{\left(\frac{M_{H_2}}{M_*} + \frac{M_{HI}}{M_*} + 1\right)} = \frac{12\%}{12\% + 3.5\% + 1} = 10.4\% . \quad (29)$$

In both the xCOLDGASS-low and xCOLDGASS-high, the SDSS *r*-band effective radius (see the **R50KPC** columns in Tables 1 and 2) measured to the galaxy had given, and it could be assigned to the value of the *R*. Hence, we calculated the mean value on the **R50KPC** to the galaxy in the xCOLDGASS-low, i.e. 2.18 kpc, as well as to the one in the xCOLDGASS-high, i.e. 3.85 kpc.

In noticing our galaxy selection, all the star-forming galaxies in the local Universe had been selected into the xCOLDGASS-low as well as the xCOLDGASS-high. Thus, we assumed the *C* by the value of 1.36 to both the xCOLDGASS-low and xCOLDGASS-high, under the consideration of the isotropic virial estimate (Schreiber et al. 2009; Tacconi et al. 2013). Moreover, due to considering that the property of our Milky Way Galaxy might be similar to those of the local star-forming galaxies selected in our two subsamples, we assumed the $\zeta$ by the value of 4.3 (I18).

With those above values determined for the xCOLDGASS-low and xCOLDGASS-high and their LR results of Table 3, we substituted them to equation (25) and evaluated the $\psi$ value for the xCOLDGASS-low as $1.40 \pm 0.30$ and the one for the xCOLDGASS-high as $1.60 \pm 0.10$. As the initiative test, for the $\psi$ result, the value following the plus-minus sign indicates the error calculated from the square root of the sum of the quadratic $\alpha$ error and the quadratic $\beta$ error, with the two-digit manner being adopted as the presentation.

Therefore, in this subsection, there were two empirical identities defined in the two galactic $M_*$ ranges, which can be summarized as:

$$A + \alpha - \beta \times B = (1.4 \pm 0.3) \times \log_{10} \left(\frac{f_{mol} \times R \times C^2}{\zeta \times G}\right)$$
$$\text{for } 10^9 \, M_\odot < M_* < 10^{10} \, M_\odot , \quad (30)$$

and

$$A + \alpha - \beta \times B = (1.6 \pm 0.1) \times \log_{10} \left(\frac{f_{mol} \times R \times C^2}{\zeta \times G}\right)$$
$$\text{for } 10^{10} \, M_\odot < M_* < 10^{11.5} \, M_\odot . \quad (31)$$

## 5 CONCLUSION

Following the end in Section 4.1, as indicated by the different values on the $\psi$, two distinctive empirical identities occurring in two different galactic $M_*$ ranges were manifested in appearance, however, while observing their uncertainties of the $\psi$ values, the identity in the higher $M_*$ range seems to be consistent to the one in the lower $M_*$ range in a $2\sigma$ level.

As our initial test on the evaluation of the $\psi$ value, the uncertainties of the $\alpha$ and $\beta$ were included without any involvement of the uncertainties of the $f_{mol}$, $R$, $C$, and $\zeta$, since we were believing in the well measurement done in the sample data base and literature data and considering the existence of the CO LFR as a priority in our LR analysis. The $\alpha$ and $\beta$ results were measured based on the valid existence of CO LFR in Section 4, therefore, we conclude that the $2\sigma$ deviation occurring in those 'true' intercepts of the CO LFRs was conveyed via the $\psi$ evaluation and manifested on that $2\sigma$-level consistency occurring in between equations (30) and (31). Then, by the procedure of that $\psi$ evaluation, it is showing that the parameter $\psi$ itself, used to link the statistical parameter to the physical parameter, can also occur validly.

At least, through our study, the CO LFR was confirmed in valid with the local star-forming galaxy sample which had been selected by our stringent requirement stated in Section 3.1. And, further, we also corroborated that the relationship between the mean $f_{mol}$ and verified CO LFR of the galaxy sample can be stated via the empirical identity proposed in equation (24).

Furthermore, as a thought, equation (30) or (31) (or, generally, equation 24) may imply an application on the estimation of the galaxy property (i.e. the mass fraction, mass-to-light ratio, and physical size) to a newly observed local star-forming galaxy.

Issues may be expectedly addressed that whether or not the form of the identity in equation (24) still occurs as the data of the star-forming galaxy lying in a higher redshift (e.g. $z > 0.05$) are being utilized, and, how do we state a relationship between the statistical property of a verified CO LFR and a mean galaxy property while the datum of a AGN sample is in use? To understand these issues, as a future work, a more effort will be taken to construct our data base compiling a different-type galaxy observed in a distinct redshift range.

## ACKNOWLEDGEMENTS

We sincerely and greatly appreciate the referee's reviewing and helpful suggestion. Y-HW and J-FW acknowledge the NSFC (National Natural Science Foundation of China) grants 12033004, 12221003, and 12333002. C-WT was supported by the NSFC grant 11988101. Y-HW acknowledges the grant of the 2021 International Postdoctoral Exchange Fellowship Program (Talent Introduction Program, Grant Number: YJ2021031) and the resources in the Xiamen University of China. Y-HW acknowledges the grant of the Xinjiang Tianchi Talent Program in 2024. Y-HW is grateful to three graduated students, Kun-Peng Shi, Lin Zhu, and Wen-Yu Zhong, for their works on compiling and selecting the galaxy data. Y-HW is grateful to the Grammarly (https://www.grammarly.com) and an on-line English assist website (https://academic.chatwithpaper.org) for the English correction on the grammatical and wording issues of this article.

This paper has been typeset from a TEX/LATEX file prepared by the author.